\documentclass[aps,prl,twocolumn,groupedaddress,showpacs,a4paper]{revtex4}
\usepackage{graphicx}

\begin{document}


\title{Dynamics of Quantum Dot Nuclear Spin Polarization Controlled by a Single Electron}



\author{P. Maletinsky}
\email[]{patrickm@phys.ethz.ch}
\author{A. Badolato}
\author{A. Imamoglu}
\affiliation{Institute of Quantum Electronics, ETH-H\"{o}nggerberg,
CH-8093, Z\"{u}rich, Switzerland}

\date{\today}

\begin{abstract}
We present an experimental study of the dynamics underlying the buildup and decay of dynamical nuclear spin polarization in a single semiconductor quantum dot. Our experiment shows that the nuclei can be polarized on a time scale of a few milliseconds, while their decay dynamics depends drastically on external parameters. We show that a single electron can very efficiently depolarize the nuclear spins and discuss two processes that can cause this depolarization. Conversely, in the absence of a quantum dot electron, the lifetime of nuclear spin polarization is on the time scale of a second, most likely limited by the non-secular terms of the nuclear dipole-dipole interaction. We can further suppress this depolarization rate by $1-2$ orders of magnitude by applying an external magnetic field exceeding $1~$mT.
\end{abstract}


\maketitle


Optically active, self assembled single quantum dots (QDs) present an excellent system for studying optically induced dynamical nuclear spin polarization (DNSP) on an isolated ensemble of $\sim10^{4-5}$ nuclear spins. While the dynamics of DNSP of nuclei close to paramagnetic impurities in bulk semiconductors has already been studied \cite{Paget1982}, addressing a single, isolated island of spin polarized nuclei has not been possible up to now. Studying the dynamics of DNSP in a single QD has the advantage of removing effects of sample inhomogeneities and crosstalk between the individual islands of spin polarized nuclei. Also, the different atomic composition and strain distribution of the QD compared to its surrounding host material further decouples the QD nuclear spins from their environment. These facts distinguish the coupled QD electron-nuclear spin system as a well isolated system of a single electron spin, coupled to a slowly varying, small nuclear spin reservoir. A further interesting aspect of this system is its similarity to the Jaynes-Cummings model in quantum optics \cite{Christ2006}, with the fully polarized nuclear spin state corresponding to the cavity vacuum state. Controlling and understanding this system to a higher degree might lead to interesting experiments such as the coherent exchange of information between the electron and the nuclear spin reservoir \cite{Taylor2004,Song2005}.

Optical orientation of QD nuclear spins has experimentally been demonstrated by a few groups \cite{Gammon1997,Brown1996,Eble2005,Lai2006,Tartakovskii2007}. However, the degree of DNSP achieved in these experiments has been limited to $\sim10-20$ percent. A detailed analysis of the formation as well as of the limiting factors of DNSP is thus required and might open ways to reach higher degrees of DNSP. A key ingredient for this understanding is the knowledge of the relevant timescales of the dynamics of nuclear spin polarization. Many questions like the respective roles of nuclear spin diffusion, quadrupolar relaxation and trapped excess QD charges on the depolarization of the nuclear spin system remain open up to now. While the buildup time of DNSP ($\tau_{\rm buildup}$) is likely to be dependant on the way the nuclear spin system is addressed, the DNSP decay time ($\tau_{\rm decay}$) is an inherent property of the isolated nuclear spin system of a QD. Experimental determination of $\tau_{\rm decay}$, which directly yields the correlation time of the fluctuations of the nuclear spin projection along the axis in which the nuclei are polarized \cite{Coish2007}, is crucial for understanding the limits of electron spin coherence in QDs \cite{MerkulovRosen2002}.

In this work, we investigate the dynamics of DNSP in an individual, self-assembled InGaAs QD at $T=5~$K. Photoluminescence (PL) of the negatively charged exciton ($X^{-1}$) is studied under resonant excitation in one of the excited QD states. It has been shown previously that under the appropriate excitation conditions, the QD nuclear spins can be polarized to a degree of $\sim15\%$. DNSP can then be measured through the Zeeman splitting of the $X^{-1}$ recombination line in the resulting nuclear magnetic field \cite{Lai2006}. This energy shift due to the spin polarized nuclei is commonly referred to as the Overhauser shift (OS). We studied the dynamics of DNSP, both, at zero magnetic field as well as in the presence of an external magnetic field of magnitude $\sim220~$mT. 

The sample was grown by molecular beam epitaxy on a $(100)$ semi-insulating GaAs substrate. The InGaAs QDs are spaced by $25~$nm of GaAs from a doped n$^{++}$-GaAs layer, followed by $30~$nm of GaAs and 29 periods of an AlAs/GaAs ($2/2~$nm) superlattice barrier which is capped by $4$-nm of GaAs. A bias voltage is applied between the top Schottky and back Ohmic contacts to control the charging state of the QD. Spectral features presented in this work were obtained at the center of the $X^{-1}$ stability plateau in gate voltage, where PL counts as well as the resulting OS were maximized \cite{Lai2006}. The low density of QDs ($<0.1~\mu m^{-2}$) allows us to address a single QD using the micro-photoluminescence ($\mu$-PL) setup described in more detail in \cite{Lai2006}. The spectral resolution of the system is determined by the spectrometers charge coupled device (CCD) pixel separation and amounts to $\sim30~\mu$eV. However, the precision to which the emission energy of a given spectral line can be determined, can be increased to $\sim2\mu$eV, by calculating a weighted average of the emission energy over the relevant CCD pixels \cite{Maletinsky2007}.

We use a ``pump-probe'' technique to investigate the dynamics of buildup and decay of DNSP. An acousto-optical modulator (AOM) serves as a fast switch of excitation light intensity, producing light pulses of variable lengths, with rise- and fall-times of $\sim600~$ns. We differentiate between ``pump'' pulses of duration $\tau_{\rm pump}$, used to polarize the nuclear spins, followed by ``probe'' pulses of length $\tau_{\rm probe}$, used to measure the resulting degree of DNSP. The intensity of each pulse corresponds to the saturation intensity of the observed emission line, maximizing both, the resulting OS and the signal to noise ratio (SNR) of the measurement. A mechanical shutter placed in the PL collection path is used to block the pump pulses, while allowing the probe pulses to reach the spectrometer. Pump and probe pulses are separated by a waiting time $\tau_{\rm wait}$ with a minimal length of $0.5~$ms, limited by the jitter of the mechanical shutter opening time. In order to measure the buildup (decay) time of DNSP, $\tau_{\rm pump}$ ($\tau_{\rm wait}$) are varied respectively while keeping all other parameters fixed. The timing and synchronization of the individual pulses is computer controlled via a digital acquisition card operating at a clock period of $2~\mu$s, which sets the time resolution of the pulse sequences. Individual pump-probe sequences are repeated and the corresponding probe pulses are accumulated on the spectrometer CCD in order to obtain a reasonable SNR. We verify a posteriori that individual pump-probe pairs are separated by much more than the DNSP decay time.

\begin{figure}
\includegraphics[width=\columnwidth]{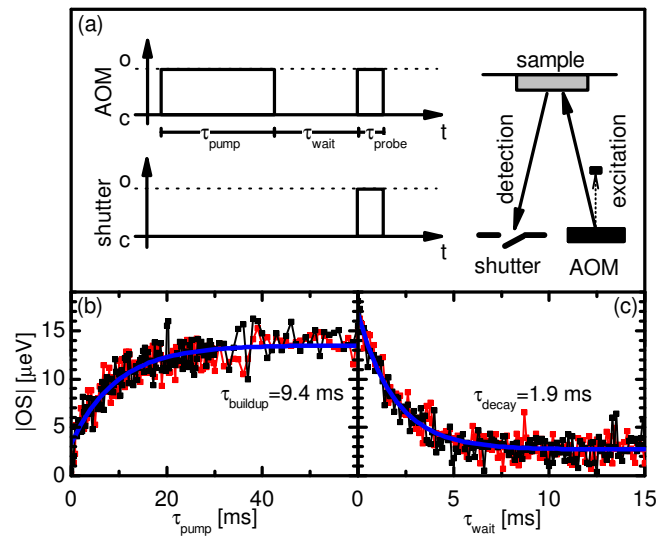} 
\caption{\label{FigUpDown} (Color online). (a) Schematic of the pulse sequences used in the buildup and decay time measurements of DNSP. An acousto optical modulator (AOM) deflects the excitation beam on and off the sample, serving as a fast switch (o (c) denote the open (closed) state, respectively). The AOM creates pump (probe) pulses of respective lengths $\tau_{\rm pump}$ ($\tau_{\rm probe}$), separated by a waiting time $\tau_{\rm wait}$. A mechanical shutter blocks the pump pulse from reaching the spectrometer, while letting the probe pulse pass. (b) DNSP buildup curves obtained by varying $\tau_{\rm pump}$ at fixed $\tau_{\rm wait}$ ($0.5~$ms) and $\tau_{\rm probe}$ ($0.2~$ms). The red (black) data points correspond to QD excitation with light of positive (negative) helicity. The blue line is an exponential fit, yielding a buildup time of $\tau_{\rm buildup}=9.4~$ms. (c) DNSP decay curves obtained by varying $\tau_{\rm wait}$, at fixed $\tau_{\rm pump}$ ($50~$ms) and $\tau_{\rm probe}$ ($0.5~$ms). The color coding is identical to (a). The exponential fit reveals a decay time of $\tau_{\rm decay}=1.9~$ms.}
\end{figure}

Figure \ref{FigUpDown}(b) and (c) show the results for buildup and decay curves of DNSP obtained with this technique. The resulting curves fit surprisingly well to a simple exponential, yielding 
$\tau_{\rm buildup}=9.4~$ms and $\tau_{\rm decay}=1.9~$ms \footnote{A simple rate equation model predicts deviations from an exponential dependance due to the feedback of DNSP on the nuclear spin cooling rate \cite{ Maletinsky2007}. However, the limited SNR of our experiment and the finite length of the probe pulses do not allow us to observe these deviations.}. The small residual OS observed for $\tau_{\rm pump}=0$ ($\tau_{\rm wait}\gg\tau_{\rm decay}$) in the buildup (decay) time measurement is due to the nuclear polarization created by the probe pulse. Comparing our experimental findings to previous experiments is not straightforward since, to the best of our knowledge, the dynamics of DNSP without an applied magnetic field has not been studied up to now. However, in experiments performed at external magnetic fields of $\sim1~$T, the buildup time of DNSP was estimated to be on the order of a few seconds \cite{Gammon2001,Maletinsky2007}. Also, previous experimental results in similar systems revealed DNSP decay times on the order of minutes \cite{Paget1982}. It is thus at first sight surprising that we find a DNSP decay time as short as a few milliseconds.

\begin{figure}
\includegraphics[width=\columnwidth]{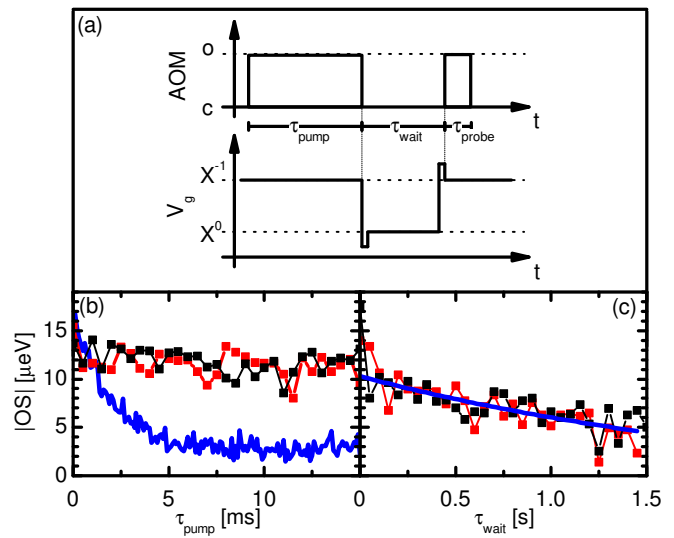}
\caption{\label{FigVgSwitch} (Color online). (a) Timing diagram for the gate voltage switching experiment: During the period $\tau_{\rm wait}$, the QD gate voltage is switched to a value where the neutral exciton is the stable QD charge complex. Using transient pulses, the switching time is $30~\mu$s. Ejecting the residual QD electron removes its effect on DNSP depolarization. This is demonstrated in (b), which shows DNSP decay time measurements in the absence of the residual QD electron. The red (black) data points represent DNSP decay under $\sigma^+$ ($\sigma^-$) excitation. For comparison, the blue curve shows the mean of the data presented in Fig. \ref{FigUpDown}(b). (c) Same measurement as in (b), but over a longer timescale. The exponential fit (blue) indicates a decay time constant of $\tau_{\rm decay}\sim2.3~$s.}
\end{figure}

A possible cause for the fast decay of DNSP is the presence of the residual QD electron even in the absence of optical pumping. We study its influence on $\tau_{\rm decay}$ with the following experiment: While the nuclear spin polarization is left to decay, we apply a voltage pulse to the QD gate electrodes, ejecting the residual electron from the QD into the nearby electron reservoir. This is achieved by switching the QD gate voltage to a value where the dominant spectral feature observed in PL stems from the recombination of the neutral exciton ($X^{0}$). Using transient voltage pulses, we are able to perform this ``gate voltage switching'' on a timescale of $30~\mu$s. Before sending the probe pulse onto the QD, the gate voltage is switched back to its initial value in order to collect PL from $X^{-1}$ recombination. The dramatic effect of this gate voltage pulsing on DNSP lifetime is shown in Fig. \ref{FigVgSwitch}(b). On the timescale of the previous measurements, almost no DNSP decay can be observed anymore. By prolonging $\tau_{\rm wait}$ up to a few seconds (Fig. \ref{FigVgSwitch}(c)), DNSP decay of the unperturbed nuclear system can be measured to be $\tau_{\rm decay}\sim2.3~$s. We note that the increase of $\tau_{\rm wait}$ necessary for this experiment results in a reduced SNR, which makes an exact determination of $\tau_{\rm decay}$ difficult. 

The role of the residual electron in depolarizing the nuclear spins was further confirmed in two independent measurements (not shown here). First, we perform a modified version of the gate voltage switching experiment: During the interval $\tau_{\rm wait}$, the gate voltage is switched to a regime where the QD ground state consists of two electrons in a spin singlet state \cite{Urbaszek2003}. This state doesn't couple to the nuclear spins and the measured $\tau_{\rm decay}$ is again on the order of seconds. The second control experiments consists in measuring DNSP dynamics at a constant gate voltage where the positively charged exciton ($X^{+1}$) is the stable QD charge complex. $X^{+1}$ has previously been shown to lead to DNSP \cite{Lai2006}. However in this case, no electron is left in the QD after exciton recombination and the corresponding DNSP decay channel is not present. As expected, $\tau_{\rm decay}$ is also on the order of seconds for this case.

We argue that two mechanisms could lead to the efficient decay of DNSP due to the residual electron. First, the presence of a QD conduction band electron leads to indirect coupling of nuclear spins in the QD \cite{Abragam1961}. The resulting rate of nuclear spin depolarization has been estimated to be on the order of $T_{\rm ind}^{-1}\sim A^2/N^{3/2}\Omega_e$ \cite{Klauser2006}. Here, $A$ is the hyperfine coupling constant ($\sim100~\mu$eV), $N\sim10^{4-5}$ the number of nuclei in the QD and $\Omega_e$ the electron spin splitting.
In order to get a rough estimate of the resulting timescale, we take $\Omega_e$ to be constant and equal to half the maximum measured OS, despite the fact that $\Omega_e$ actually varies during the course of nuclear spin depolarization. With these values, we obtain a nuclear spin depolarization time of a few $\mu$s. This is an upper bound for the corresponding DNSP decay rate which will be slowed by additional effects like the inhomogeneous Knight field the nuclei are exposed to. Secondly, the spin of the residual electron is randomized due to co-tunnelling to the close-by electron reservoir on a timescale of $\tau_{\rm el}\sim20~$ns \cite{Smith2005}. This electron spin depolarization is then mapped onto the nuclear spin system via hyperfine flip-flop events. Taking into account the detuning $\Omega_e$ of the two electron spin levels, the resulting nuclear spin depolarization rate can be estimated to be $T_{\rm 1e}^{-1}\sim(A/N\hbar)^2 / \Omega_e^2 \tau_{\rm el}$ \cite{Meier1984}, which is on the order of a hundred ms for the same parameters as before.

\begin{figure}
\includegraphics[width=\columnwidth]{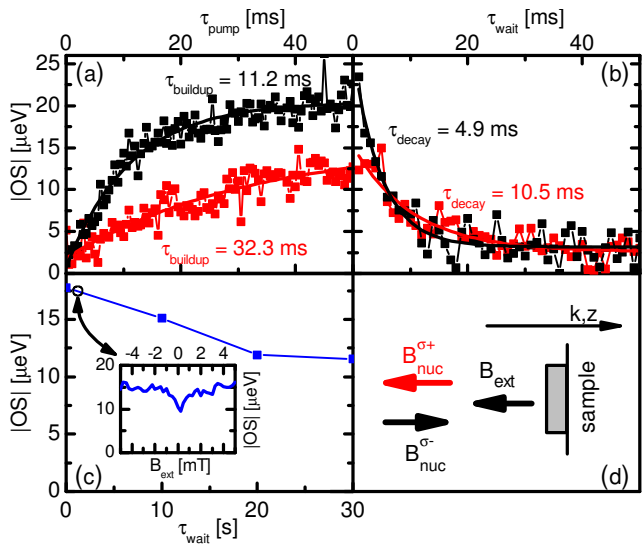}
\caption{\label{FigBext} (Color online). Measurements of buildup and decay of DNSP in an external magnetic field $B_{\rm ext}\sim -220~$mT: (a) Buildup of DNSP. In the presence of $B_{\rm ext}$, it is more efficient and thus faster to produce a nuclear magnetic field compensating the latter (black, $\sigma^-$ excitation) than one that enforces it (red, $\sigma^+$ excitation) \cite{Maletinsky2007}. (b) If DNSP decay is mediated through the residual QD electron, it is again more efficient to depolarize the nuclei if the total effective magnetic field seen by the electron is minimized. The color coding is the same as in (a). Solid curves in (a) and (b) show exponential fits to the data, the resulting buildup- and decay times are given in the figures. (c) Decay of DNSP in the absence of the QD electron. Compared to the zero-field case (Fig.\ref{FigVgSwitch}c), DNSP decay time is prolonged to $\tau_{\rm decay}\sim60~$s. The inset shows OS after a waiting time of $1s$ as a function of external magnetic field. DNSP decay is suppressed on a magnetic field scale of $\sim1~$mT, indicative of DNSP decay mediated by nuclear dipole-dipole interactions. (d) shows the respective directions of the external magnetic field and the nuclear fields $B_{\rm nuc}^{\sigma^+}$ ($B_{\rm nuc}^{\sigma^-}$) induced by QD excitation with $\sigma^+$ ($\sigma^-$) polarized light.} 
\end{figure}

Our study of DNSP timescales was complemented by adding a permanent magnet to our sample. The resulting magnetic field is antiparallel to the excitation beam direction and has a magnitude of $B_{\rm ext}=-220~$mT at the site of the QD \footnote{Sign and magnitude of $B_{\rm ext}$ were determined by comparing the observed PL line splittings with the ones obtained in \cite{Maletinsky2007}, where a calibrated superconducting magnet was used.}. The buildup and decay time measurements in the presence of $B_{\rm ext}$ are shown in Fig. \ref{FigBext}. An asymmetry between the cases of $\sigma^+$ and $\sigma^-$ excitation can be observed. Exciting the QD with $\sigma^{(+)-}$-polarized light creates a nuclear field ($B_{\rm nuc}^{\sigma^{(+)-}}$) aligned (anti-)parallel to $B_{\rm ext}$. The two nuclear fields $B_{\rm nuc}^{\sigma^+}$ and $B_{\rm nuc}^{\sigma^-}$ differ in magnitude due to the dependance of the electron-nuclear spin flip-flop rate on electron Zeeman splitting \cite{Braun2006,Maletinsky2007,Tartakovskii2007}. This feedback of DNSP on the nuclear spin cooling rate makes it faster and thus more efficient to create a nuclear field that compensates $B_{\rm ext}$. Conversely, creating a nuclear field that enforces $B_{\rm ext}$, slows down nuclear spin cooling and leads to a smaller degree of DNSP. The measurements presented in Fig. \ref{FigBext} (a) and (b) confirm this picture. Since both, $\tau_{\rm buildup}$ and $\tau_{\rm decay}$ are mediated by the hyperfine flip-flop interaction, the corresponding timescales should depend on the helicity of the excitation light and thus on the direction of the resulting $B_{\rm nuc}$. Indeed, we find that $\tau_{\rm buildup}$ and $\tau_{\rm decay}$ are both increased by a factor of $\sim2-3$, when changing the polarization of the excitation light from $\sigma^-$ to $\sigma^+$.

We again performed the ``gate voltage switching'' experiment in the presence of $B_{\rm ext}$ (Fig. \ref{FigBext} (c)). Since in this case DNSP decay is not mediated by the residual QD electron, no dependance of $\tau_{\rm decay}$ on excitation light helicity was found and only the average between the two data sets ($\sigma^+$ and $\sigma^-$ excitation) is shown. Compared to the case of zero external magnetic field, the decay of nuclear spin polarization is further suppressed. Even though extracting exact numbers is difficult in this case due to the required long waiting times, we estimate $\tau_{\rm decay}$ to be on the order of a minute. This further suppression of DNSP decay rate can be induced with a magnetic field as small as $\sim1~$mT as shown in the inset of Fig. \ref{FigBext} (c): Keeping $\tau_{\rm wait}=1~$s fixed, we sweep an external magnetic field while measuring the remaining OS. The resulting dip around $B_{\rm ext}=0$ has a width of $\sim 1~$mT. This indicates that nuclear spin depolarization at zero magnetic field is governed by the non-secular terms of the nuclear dipole-dipole interactions \cite{Abragam1961}. These terms, which don't conserve angular momentum, are very effective in depolarizing nuclear spins as long as their Zeeman splitting is not much larger than the nuclear dipole-dipole energy, which corresponds to a local magnetic field $B_{L}\sim0.1~$mT \cite{Meier1984}.

Finally, we investigated the possible role of nuclear spin diffusion and the resulting DNSP of the bulk nuclei surrounding the QD. For this, we studied the dependance of $\tau_{\rm decay}$ on the nuclear spin pumping time $\tau_{\rm pump}$ for $\tau_{\rm pump}\gg\tau_{\rm buildup}$. A nuclear spin polarization in the surrounding of the QD would lead to an increase of $\tau_{\rm decay}$ with increasing $\tau_{\rm pump}$\cite{Paget1982}. However, within the experimental parameters currently accessible in our experiment, we were not able to see such a prolongation and hence any effects of polarization of the surrounding bulk nuclei. We interpret this fact as a strong indication that we indeed create and observe a very isolated system of spin polarized nuclei.

The present study of the dynamics of the QD electron-nuclear spin system revealed a surprisingly short decay time of DNSP. We were able to assign this to the role the residual QD electron plays in depolarizing the nuclear spins and propose two distinct physical mechanisms that can cause this decay: Indirect interaction of the nuclear spins as well as co-tunnelling mediated electron spin depolarization. While distinguishing these two effects is not possible with the data at hand, a systematic study of $\tau_{\rm decay}$ as a function of external magnetic fields could clarify the exact nature of DNSP decay due to the different functional dependencies of $T_{\rm 1e}$ and $T_{\rm ind}$ on $\Omega_e$. A study of DNSP dynamics as a function of external magnetic field in the absence of the residual QD electron could also clarify the role of quadrupolar interactions in DNSP. Suppressing these interactions at high magnetic fields could further increase DNSP lifetime up to several minutes. Another interesting regime for performing these studies is at magnetic fields, where the coupled electron-nuclear spin system exhibits a bistable behavior and the dynamics become highly nonlinear \cite{Braun2006,Maletinsky2007,Tartakovskii2007}.

\begin{acknowledgments}

P.M. would like to thank C.W. Lai for his great support in the lab which made this work possible. Furthermore, we would like to acknowledge W.A. Coish for fruitful discussions and thank J. Dreiser, S. F\"alt and B. Demaurex for assistance in the lab and help with sample preparation. This work is supported by NCCR-Nanoscience.
\end{acknowledgments}

\bibliography{BibNuclearSpin}

\begin{thebibliography}{19}
\expandafter\ifx\csname natexlab\endcsname\relax\def\natexlab#1{#1}\fi
\expandafter\ifx\csname bibnamefont\endcsname\relax
  \def\bibnamefont#1{#1}\fi
\expandafter\ifx\csname bibfnamefont\endcsname\relax
  \def\bibfnamefont#1{#1}\fi
\expandafter\ifx\csname citenamefont\endcsname\relax
  \def\citenamefont#1{#1}\fi
\expandafter\ifx\csname url\endcsname\relax
  \def\url#1{\texttt{#1}}\fi
\expandafter\ifx\csname urlprefix\endcsname\relax\def\urlprefix{URL }\fi
\providecommand{\bibinfo}[2]{#2}
\providecommand{\eprint}[2][]{\url{#2}}

\bibitem[{\citenamefont{Paget}(1982)}]{Paget1982}
\bibinfo{author}{\bibfnamefont{D.}~\bibnamefont{Paget}},
  \bibinfo{journal}{Phys. Rev. B} \textbf{\bibinfo{volume}{25}},
  \bibinfo{pages}{4444} (\bibinfo{year}{1982}).

\bibitem[{\citenamefont{Christ et~al.}(2006)\citenamefont{Christ, Cirac, and
  Giedke}}]{Christ2006}
\bibinfo{author}{\bibfnamefont{H.}~\bibnamefont{Christ}},
  \bibinfo{author}{\bibfnamefont{J.}~\bibnamefont{Cirac}}, \bibnamefont{and}
  \bibinfo{author}{\bibfnamefont{G.}~\bibnamefont{Giedke}}
  (\bibinfo{year}{2006}), \eprint{cond-mat/0611438}.

\bibitem[{\citenamefont{Taylor et~al.}(2004)\citenamefont{Taylor, Giedke,
  Christ, Paredes, Cirac, Zoller, Lukin, and Imamoglu}}]{Taylor2004}
\bibinfo{author}{\bibfnamefont{J.}~\bibnamefont{Taylor}},
  \bibinfo{author}{\bibfnamefont{G.}~\bibnamefont{Giedke}},
  \bibinfo{author}{\bibfnamefont{H.}~\bibnamefont{Christ}},
  \bibinfo{author}{\bibfnamefont{B.}~\bibnamefont{Paredes}},
  \bibinfo{author}{\bibfnamefont{J.}~\bibnamefont{Cirac}},
  \bibinfo{author}{\bibfnamefont{P.}~\bibnamefont{Zoller}},
  \bibinfo{author}{\bibfnamefont{M.}~\bibnamefont{Lukin}}, \bibnamefont{and}
  \bibinfo{author}{\bibfnamefont{A.}~\bibnamefont{Imamoglu}}
  (\bibinfo{year}{2004}), \eprint{cond-mat/0407640}.

\bibitem[{\citenamefont{Song et~al.}(2004)\citenamefont{Song, Zhang, Shi, and
  Sun}}]{Song2005}
\bibinfo{author}{\bibfnamefont{Z.}~\bibnamefont{Song}},
  \bibinfo{author}{\bibfnamefont{P.}~\bibnamefont{Zhang}},
  \bibinfo{author}{\bibfnamefont{T.}~\bibnamefont{Shi}}, \bibnamefont{and}
  \bibinfo{author}{\bibfnamefont{C.-P.} \bibnamefont{Sun}}
  (\bibinfo{year}{2004}), \eprint{cond-mat/0409185}.

\bibitem[{\citenamefont{Gammon et~al.}(1997)\citenamefont{Gammon, Brown, Snow,
  Kennedy, Katzer, and Park}}]{Gammon1997}
\bibinfo{author}{\bibfnamefont{D.}~\bibnamefont{Gammon}},
  \bibinfo{author}{\bibfnamefont{S.~W.} \bibnamefont{Brown}},
  \bibinfo{author}{\bibfnamefont{E.~S.} \bibnamefont{Snow}},
  \bibinfo{author}{\bibfnamefont{T.~A.} \bibnamefont{Kennedy}},
  \bibinfo{author}{\bibfnamefont{D.~S.} \bibnamefont{Katzer}},
  \bibnamefont{and} \bibinfo{author}{\bibfnamefont{D.}~\bibnamefont{Park}},
  \bibinfo{journal}{Science} \textbf{\bibinfo{volume}{277}},
  \bibinfo{pages}{85} (\bibinfo{year}{1997}).

\bibitem[{\citenamefont{Brown et~al.}(1996)\citenamefont{Brown, Kennedy,
  Gammon, and Snow}}]{Brown1996}
\bibinfo{author}{\bibfnamefont{S.~W.} \bibnamefont{Brown}},
  \bibinfo{author}{\bibfnamefont{T.~A.} \bibnamefont{Kennedy}},
  \bibinfo{author}{\bibfnamefont{D.}~\bibnamefont{Gammon}}, \bibnamefont{and}
  \bibinfo{author}{\bibfnamefont{E.~S.} \bibnamefont{Snow}},
  \bibinfo{journal}{Phys. Rev. B} \textbf{\bibinfo{volume}{54}},
  \bibinfo{pages}{17339} (\bibinfo{year}{1996}).

\bibitem[{\citenamefont{Eble et~al.}(2005)\citenamefont{Eble, Krebs, Lemaitre,
  Kowalik, Kudelski, Voisin, Urbaszek, Marie, and Amand}}]{Eble2005}
\bibinfo{author}{\bibfnamefont{B.}~\bibnamefont{Eble}},
  \bibinfo{author}{\bibfnamefont{O.}~\bibnamefont{Krebs}},
  \bibinfo{author}{\bibfnamefont{A.}~\bibnamefont{Lemaitre}},
  \bibinfo{author}{\bibfnamefont{K.}~\bibnamefont{Kowalik}},
  \bibinfo{author}{\bibfnamefont{A.}~\bibnamefont{Kudelski}},
  \bibinfo{author}{\bibfnamefont{P.}~\bibnamefont{Voisin}},
  \bibinfo{author}{\bibfnamefont{B.}~\bibnamefont{Urbaszek}},
  \bibinfo{author}{\bibfnamefont{X.}~\bibnamefont{Marie}}, \bibnamefont{and}
  \bibinfo{author}{\bibfnamefont{T.}~\bibnamefont{Amand}}
  (\bibinfo{year}{2005}), \eprint{cond-mat/0508281}.

\bibitem[{\citenamefont{Lai et~al.}(2006)\citenamefont{Lai, Maletinsky,
  Badolato, and Imamoglu}}]{Lai2006}
\bibinfo{author}{\bibfnamefont{C.~W.} \bibnamefont{Lai}},
  \bibinfo{author}{\bibfnamefont{P.}~\bibnamefont{Maletinsky}},
  \bibinfo{author}{\bibfnamefont{A.}~\bibnamefont{Badolato}}, \bibnamefont{and}
  \bibinfo{author}{\bibfnamefont{A.}~\bibnamefont{Imamoglu}},
  \bibinfo{journal}{Phys. Rev. Lett.} \textbf{\bibinfo{volume}{96}},
  \bibinfo{pages}{167403} (\bibinfo{year}{2006}).

\bibitem[{\citenamefont{Tartakovskii et~al.}(2007)\citenamefont{Tartakovskii,
  Wright, Russell, Fal'ko, Van'kov, Skiba-Szymanska, Drouzas, Kolodka,
  Skolnick, Fry et~al.}}]{Tartakovskii2007}
\bibinfo{author}{\bibfnamefont{A.~I.} \bibnamefont{Tartakovskii}},
  \bibinfo{author}{\bibfnamefont{T.}~\bibnamefont{Wright}},
  \bibinfo{author}{\bibfnamefont{A.}~\bibnamefont{Russell}},
  \bibinfo{author}{\bibfnamefont{V.~I.} \bibnamefont{Fal'ko}},
  \bibinfo{author}{\bibfnamefont{A.~B.} \bibnamefont{Van'kov}},
  \bibinfo{author}{\bibfnamefont{J.}~\bibnamefont{Skiba-Szymanska}},
  \bibinfo{author}{\bibfnamefont{I.}~\bibnamefont{Drouzas}},
  \bibinfo{author}{\bibfnamefont{R.~S.} \bibnamefont{Kolodka}},
  \bibinfo{author}{\bibfnamefont{M.~S.} \bibnamefont{Skolnick}},
  \bibinfo{author}{\bibfnamefont{P.}~\bibnamefont{Fry}}, \bibnamefont{et~al.},
  \bibinfo{journal}{Phys. Rev. Lett.} \textbf{\bibinfo{volume}{98}},
  \bibinfo{pages}{26806} (\bibinfo{year}{2007}).

\bibitem[{\citenamefont{Coish}(2007)}]{Coish2007}
\bibinfo{author}{\bibfnamefont{W.~A.} \bibnamefont{Coish}},
  \bibinfo{howpublished}{private communication} (\bibinfo{year}{2007}).

\bibitem[{\citenamefont{Merkulov et~al.}(2002)\citenamefont{Merkulov, Efros,
  and Rosen}}]{MerkulovRosen2002}
\bibinfo{author}{\bibfnamefont{I.~A.} \bibnamefont{Merkulov}},
  \bibinfo{author}{\bibfnamefont{A.~L.} \bibnamefont{Efros}}, \bibnamefont{and}
  \bibinfo{author}{\bibfnamefont{M.}~\bibnamefont{Rosen}},
  \bibinfo{journal}{Phys. Rev. B} \textbf{\bibinfo{volume}{65}},
  \bibinfo{pages}{205309} (\bibinfo{year}{2002}).

\bibitem[{\citenamefont{Maletinsky et~al.}(2007)\citenamefont{Maletinsky, Lai,
  Badolato, and Imamoglu}}]{Maletinsky2007}
\bibinfo{author}{\bibfnamefont{P.}~\bibnamefont{Maletinsky}},
  \bibinfo{author}{\bibfnamefont{C.~W.} \bibnamefont{Lai}},
  \bibinfo{author}{\bibfnamefont{A.}~\bibnamefont{Badolato}}, \bibnamefont{and}
  \bibinfo{author}{\bibfnamefont{A.}~\bibnamefont{Imamoglu}},
  \bibinfo{journal}{Phys. Rev. B} \textbf{\bibinfo{volume}{75}},
  \bibinfo{pages}{35409} (\bibinfo{year}{2007}).

\bibitem[{\citenamefont{Gammon et~al.}(2001)\citenamefont{Gammon, Efros,
  Kennedy, Rosen, Katzer, Park, Brown, Korenev, and Merkulov}}]{Gammon2001}
\bibinfo{author}{\bibfnamefont{D.}~\bibnamefont{Gammon}},
  \bibinfo{author}{\bibfnamefont{A.~L.} \bibnamefont{Efros}},
  \bibinfo{author}{\bibfnamefont{T.~A.} \bibnamefont{Kennedy}},
  \bibinfo{author}{\bibfnamefont{M.}~\bibnamefont{Rosen}},
  \bibinfo{author}{\bibfnamefont{D.~S.} \bibnamefont{Katzer}},
  \bibinfo{author}{\bibfnamefont{D.}~\bibnamefont{Park}},
  \bibinfo{author}{\bibfnamefont{S.~W.} \bibnamefont{Brown}},
  \bibinfo{author}{\bibfnamefont{V.~L.} \bibnamefont{Korenev}},
  \bibnamefont{and} \bibinfo{author}{\bibfnamefont{I.~A.}
  \bibnamefont{Merkulov}}, \bibinfo{journal}{Phys. Rev. Lett.}
  \textbf{\bibinfo{volume}{86}}, \bibinfo{pages}{5176} (\bibinfo{year}{2001}).

\bibitem[{\citenamefont{Urbaszek et~al.}(2003)\citenamefont{Urbaszek,
  Warburton, Karrai, Gerardot, Petroff, and Garcia}}]{Urbaszek2003}
\bibinfo{author}{\bibfnamefont{B.}~\bibnamefont{Urbaszek}},
  \bibinfo{author}{\bibfnamefont{R.~J.} \bibnamefont{Warburton}},
  \bibinfo{author}{\bibfnamefont{K.}~\bibnamefont{Karrai}},
  \bibinfo{author}{\bibfnamefont{B.~D.} \bibnamefont{Gerardot}},
  \bibinfo{author}{\bibfnamefont{P.~M.} \bibnamefont{Petroff}},
  \bibnamefont{and} \bibinfo{author}{\bibfnamefont{J.~M.}
  \bibnamefont{Garcia}}, \bibinfo{journal}{Phys. Rev. Lett.}
  \textbf{\bibinfo{volume}{90}}, \bibinfo{pages}{247403}
  (\bibinfo{year}{2003}).

\bibitem[{\citenamefont{Abragam}(1961)}]{Abragam1961}
\bibinfo{author}{\bibfnamefont{A.}~\bibnamefont{Abragam}},
  \emph{\bibinfo{title}{The principles of nuclear magnetism}}
  (\bibinfo{publisher}{Clarendon Press}, \bibinfo{address}{Oxford},
  \bibinfo{year}{1961}).

\bibitem[{\citenamefont{Klauser et~al.}(2004)\citenamefont{Klauser, Coish, and
  Loss}}]{Klauser2006}
\bibinfo{author}{\bibfnamefont{D.}~\bibnamefont{Klauser}},
  \bibinfo{author}{\bibfnamefont{W.~A.} \bibnamefont{Coish}}, \bibnamefont{and}
  \bibinfo{author}{\bibfnamefont{D.}~\bibnamefont{Loss}},
  \bibinfo{journal}{Phys. Rev. B} \textbf{\bibinfo{volume}{73}},
  \bibinfo{pages}{205302} (\bibinfo{year}{2004}).

\bibitem[{\citenamefont{Smith et~al.}(2005)\citenamefont{Smith, Dalgarno,
  Warburton, Govorov, Karrai, Gerardot, and Petroff}}]{Smith2005}
\bibinfo{author}{\bibfnamefont{J.~M.} \bibnamefont{Smith}},
  \bibinfo{author}{\bibfnamefont{P.~A.} \bibnamefont{Dalgarno}},
  \bibinfo{author}{\bibfnamefont{R.~J.} \bibnamefont{Warburton}},
  \bibinfo{author}{\bibfnamefont{A.~O.} \bibnamefont{Govorov}},
  \bibinfo{author}{\bibfnamefont{K.}~\bibnamefont{Karrai}},
  \bibinfo{author}{\bibfnamefont{B.~D.} \bibnamefont{Gerardot}},
  \bibnamefont{and} \bibinfo{author}{\bibfnamefont{P.~M.}
  \bibnamefont{Petroff}}, \bibinfo{journal}{Phys. Rev. Lett.}
  \textbf{\bibinfo{volume}{94}}, \bibinfo{pages}{197402}
  (\bibinfo{year}{2005}).

\bibitem[{\citenamefont{Meier}(1984)}]{Meier1984}
\bibinfo{author}{\bibfnamefont{F.}~\bibnamefont{Meier}},
  \emph{\bibinfo{title}{Optical orientation}}
  (\bibinfo{publisher}{North-Holland}, \bibinfo{address}{Amsterdam},
  \bibinfo{year}{1984}).

\bibitem[{\citenamefont{Braun et~al.}(2006)\citenamefont{Braun, Urbaszek,
  Amand, Marie, Krebs, Eble, Lemaitre, and Voisin}}]{Braun2006}
\bibinfo{author}{\bibfnamefont{P.-F.} \bibnamefont{Braun}},
  \bibinfo{author}{\bibfnamefont{B.}~\bibnamefont{Urbaszek}},
  \bibinfo{author}{\bibfnamefont{T.}~\bibnamefont{Amand}},
  \bibinfo{author}{\bibfnamefont{X.}~\bibnamefont{Marie}},
  \bibinfo{author}{\bibfnamefont{O.}~\bibnamefont{Krebs}},
  \bibinfo{author}{\bibfnamefont{B.}~\bibnamefont{Eble}},
  \bibinfo{author}{\bibfnamefont{A.}~\bibnamefont{Lemaitre}}, \bibnamefont{and}
  \bibinfo{author}{\bibfnamefont{P.}~\bibnamefont{Voisin}},
  \bibinfo{journal}{Phys. Rev. B} \textbf{\bibinfo{volume}{74}},
  \bibinfo{pages}{245306} (\bibinfo{year}{2006}).

\end{thebibliography}

\end{document}